\documentclass[12pt,epsf,amssymb]{article} \textheight =23 truecm
\def\lsim{\raise0.3ex\hbox{$<$\kern-0.75em\raise-1.1ex\hbox{$\sim$}}}
\def\gsim{\raise0.3ex\hbox{$>$\kern-0.75em\raise-1.1ex\hbox{$\sim$}}}
\def\noi{\noindent} \def\nn{\nonumber} \def\bea{\begin{eqnarray}}
\def\eea{\end{eqnarray}} \def\beq{\begin{equation}}
\def\eeq{\end{equation}} 
\def\beeq{\begin{eqnarray}} \def\eeeq{\end{eqnarray}} \def\R{ {\rm R
\kern -.31cm I \kern .15cm}} \def\C{ {\rm C \kern -.15cm \vrule
width.5pt \kern .12cm}} \def\Z{ {\rm Z \kern -.27cm \angle \kern
.02cm}} \def\N{ {\rm N \kern -.26cm \vrule width.4pt \kern .10cm}}
\def\1{{\rm 1\mskip-4.5mu l} }

\usepackage{graphics} \usepackage{graphicx} \usepackage{epsfig}

\begin{document} \begin{center} 

{\large \bf Bound on the curvature of the Isgur-Wise function} \\
{\large \bf of the baryon semileptonic decay $\Lambda_{\bf b} \to \Lambda_{\bf c} \ell \overline{\nu}_{\ell}$}
\par \vskip 1 truecm

 {\bf A. Le Yaouanc, L. Oliver and J.-C. Raynal}
\par \vskip 3 truemm

{\it Laboratoire de Physique Th\'eorique}\footnote{Unit\'e Mixte de
Recherche UMR 8627 - CNRS }\\    {\it Universit\'e de Paris XI,
B\^atiment 210, 91405 Orsay Cedex, France} 

\end{center}

\vskip 1 truecm

\begin{abstract} 
In the heavy quark limit of QCD, using the Operator Product Expansion, the formalism of Falk for hadrons of arbitrary spin, and the non-forward amplitude, as proposed by Uraltsev, we formulate sum rules involving the Isgur-Wise function $\xi_{\Lambda} (w)$ of the baryon transition $\Lambda_b \to \Lambda_c \ell \overline{\nu}_{\ell}$, where the light cloud has $j^P=0^+$ for both initial and final baryons. We recover the lower bound for the slope $\rho_\Lambda^2 = - \xi '_\Lambda (1) \geq 0$ obtained by Isgur et al., and we generalize it by demonstrating that the IW function $\xi_{\Lambda} (w)$ is an alternate series in powers of $(w-1)$, i.e. $(-1)^n \xi_{\Lambda}^{(n)} (1) \geq 0$. Moreover, exploiting systematically the sum rules, we get an improved lower bound for the curvature in terms of the slope, $\sigma_\Lambda^2 = \xi ''_\Lambda (1) \geq {3 \over 5} [\rho_\Lambda^2 + (\rho_\Lambda^2)^2]$. This bound constrains the shape of the Isgur-Wise function and it will be compelling in the analysis of future precise data on the differential rate of the baryon semileptonic decay $\Lambda_b \to \Lambda_c \ell \overline{\nu}_{\ell}$, that has a large measured branching ratio, of about 5 \%.
\end{abstract}

\vskip 2 truecm

\noi LPT Orsay 08-65 \par 
\noi July 2008
\par \vskip 1 truecm

\newpage \pagestyle{plain}

\section{Introduction.} \hspace*{\parindent} 
Let us first briefly review the situation of meson semileptonic decays $\overline{B} \to D^{(*)}\ell \overline{\nu}_{\ell}$, and next shift to the topic of the present paper, the baryon decay $\Lambda_b \to \Lambda_c \ell \overline{\nu}_{\ell}$. The meson case will illuminate some aspects of the baryon case. In the leading order of the heavy quark expansion of QCD, Bjorken sum rule (SR) \cite{1r,2r} relates the slope of the elastic heavy meson Isgur-Wise (IW) function $\xi (w)$, to the IW functions of the transition between the ground state $j^P = {1 \over 2}^-$ and the $j^P = {1 \over 2}^+, {3 \over 2}^+$ excited states, $\tau_{1/2}^{(n)} (w)$, $\tau_{3/2}^{(n)} (w)$, at zero recoil $w=1$ ($n$ is a radial quantum number). This SR leads to the lower bound $\rho^2 = - \xi ' (1) \geq {1\over 4}$. A new SR was formulated by Uraltsev in the heavy quark limit \cite{3r}, involving also $\tau_{1/2}^{(n)} (w)$, $\tau_{3/2}^{(n)} (w)$, that implies, combined with Bjorken SR, the much stronger lower bound
\beq
\label{1e}
\rho^2 \geq {3 \over 4}
\eeq

A basic ingredient in deriving this bound was the consideration of the non-forward amplitude $ \overline{B}(v_i) \to D^{(n)}(v') \to  \overline{B}(v_f)$, allowing for general $v_i$, $v_f$, $v'$ and where $B$ is a ground state meson. In refs. \cite{4r,5r,6r} we have developed, in the heavy quark limit of QCD, a manifestly covariant formalism within the Operator Product Expansion (OPE), using the whole tower of heavy meson states \cite{7r}. We did recover Uraltsev SR plus a general class of SR that allow to bound also higher derivatives of the IW function. In particular, we found two bounds for the curvature $\sigma^2 = \xi '' (1) $ in terms of $\rho^2$, namely
 \bea
\label{2e}
&&\sigma^2 \geq {5 \over 4} \ \rho^2 \\
&&\sigma^2 \geq {1 \over 5} \left [ 4 \rho^2 + 3(\rho^2)^2 \right ] \ .
\label{3e}
\eea
 
\noi that both reduce to
\beq
\label{4e}
\sigma^2 \geq {15 \over 16}
\eeq

\noi for the lower limit (\ref{1e}). \par

On the other hand, we found also lower bounds for all higher derivatives, namely \cite{5r}
\beq
\label{5e}
(-1)^L\xi^{(L)}(1) \geq {(2L+1)!! \over 2^{2L}}
\eeq

\noi that reduce to (\ref{1e}) and (\ref{4e}) for the slope and the curvature. \par

In the baryon ``elastic'' case, several types of transitions can be considered, namely $\Lambda_b \to \Lambda_c$, $\Sigma_b \to \Sigma_c$ (and the related transitions to the latter, namely $\Sigma_b \to \Sigma_c^*$, $\Sigma_b^* \to \Sigma_c^*$), where $\Lambda_Q$ has isospin $I = 0$ and $J^P =  {1 \over 2}^+$, and $\Sigma_Q$ has $I= 1$,  $J^P =  {1 \over 2}^+$ ($I=1$, $J^P =  {3 \over 2}^+$ for  $\Sigma_Q^*$), as well as the corresponding transitions for $\Xi_b$, $\Omega_b$, etc. \par

We will here concentrate on the semileptonic decay $\Lambda_b \to \Lambda_c \ell \overline{\nu}_{\ell}$. Some decay modes have already been mesured for the $\Lambda_b$, in particular this semileptonic decay, $BR(\Lambda_b \to \Lambda_c \ell \overline{\nu}_{\ell}) \cong 5\%$, a large fraction of the inclusive decay $BR(\Lambda_b \to \Lambda_c \ell \overline{\nu}_{\ell} + \hbox{ anything}) \cong 10\%$. Hopefully, in the near future, at the LHC-b program, the exclusive semileptonic decay $\Lambda_b \to \Lambda_c \ell \overline{\nu}_{\ell}$ will be measured in detail, in particular its differential rate. Therefore, it is of interest to study the properties of the corresponding elastic IW function, that we denote here by $\xi_{\Lambda}(w)$.\par

In the heavy quark limit for $b$ and $c$ quarks, the baryons $\Lambda_b$ and $\Lambda_c$ have a light cloud with $j = 0$. Isgur et al. \cite{8r} formulated the equivalent of Bjorken sum rule for this case,
\beq
\label{6e}
\rho^2_\Lambda = - \xi '_\Lambda (1) = \sum_{n\geq 0} \left [ \tau_1^{(n)} (1) \right ]^2
\eeq

\noi obtaining therefore
\beq
\label{7e}
\rho^2_\Lambda \geq 0
\eeq

\noi The quantities $\tau_1^{(n)} (1) $ denote the $0^+ \to 1^-$ ($j^P$ of the light cloud, $n$ denotes a radial quantum number) IW functions at zero recoil. We use this notation to keep track of the analogy with the meson case, and we will make below the link with the notation of Falk \cite{7r}. Let us still underline that only intermediate states $\Lambda_c^{(n)}$ with isospin $I=0$ can contribute.\par

Baryon semileptonic decays have aroused some interest. New symmetries in these decays were formulated by Isgur and Wise \cite{9r}. Baryons of arbitrary spin were considered in detail in \cite{7r}, whose formalism we use below. On the other hand, the IW function for $\Lambda_b \to \Lambda_c \ell \overline{\nu}_{\ell}$ was studied in the large $N_c$ limit by Jenkins et al. \cite{10r} and by Chow \cite{11r}. Power corrections to baryon form factors were studied by Georgi et al. \cite{1bisr}, Falk and Neubert \cite{12r} and Mannel and Roberts \cite{13r}. Within the dispersive approach in QCD, baryon form factors have been studied at finite mass by C. Boyd et al.  \cite{2bisr,3bisr}. The slope for the IW was computed within the QCD Sum Rules approach by Huang et al. \cite{14r}. Extensive and useful review papers on heavy baryons are those of K\"orner et al. \cite{15r} and Falk  \cite{16r}. Finally, a study of semileptonic decays of $\Lambda_b$, $\Lambda_c$ baryons in quark models has been performed by Pervin et al. \cite{17r}, where it has been pointed out that the extension to baryons of our meson bound (\ref{3e}) was lacking. The present paper answers to this need. \par

Thus, our aim is to extend the program realized in the meson case to the baryon transition $\Lambda_b \to \Lambda_c \ell \overline{\nu}_{\ell}$ and formulate the equivalent relations to the results of the meson case, in particular (\ref{3e}) and (\ref{5e}).\par

Let us underline that there is an important difference between the meson case $\overline{B} \to D^{(*)}$ and the baryon case $\Lambda_b \to \Lambda_c$. In the sum rules (SR) of the meson case we had contributions where the light cloud has, for a given orbital angular momentum $L$, two possible values $j^P = \left ( L \pm {1 \over 2} \right )^P$, $P = (-1)^{L+1}$, where $S = {1 \over 2}$ stands for the spin of the spectator light quark. In the baryon transition $\Lambda_b  \to \Lambda_c$, since the two spectator quarks have total spin and isospin $S = I = 0$, we have, for a given $L$, a single type of intermediate states, with $j^P = L^P$, $P = (-1)^L$.\par
There is another important difference between the meson and the baryon cases. This concerns the $1/m_Q$ corrections. Georgi et al.  \cite{1bisr} have demonstrated that the first order $1/m_Q$ corrections in $\Lambda_b \to \Lambda_c \ell \overline{\nu}_{\ell}$ are given simply in terms of the IW function $\xi_{\Lambda}(w)$ and the constant $\overline{\Lambda}_{\Lambda} = m_{\Lambda_b}-m_b$. This is to be distinguished from the meson case, where the $1/m_Q$ corrections depend on the IW function $\xi(w)$, the constant $\overline{\Lambda} = m_B-m_b$ and on another function $\xi_3(w)$, as shown for example by Falk and Neubert \cite{4bisr}. Thus, the decay $\Lambda_b \to \Lambda_c \ell \overline{\nu}_{\ell}$ has $1/m_Q$ corrections much better controlled than the meson case.\par 
The paper is organized as follows. In Section~2 we recall the SR for mesons within the framework of the non-forward amplitude. In Section~3 we formulate the SR in the baryon case and underline the differences with the meson case.
In Section~4 we generalize the inequality (\ref{7e}) for all the derivatives, demonstrating that the baryon IW function $ \xi_{\Lambda}(w)$ is an alternate series in powers of $(w-1)$. In Section~5, exploiting systematically all the SR that can be obtained for the baryon case, we get an improved bound for the curvature of the IW function, that is the equivalent of the inequality (\ref{3e}) for the meson case. In Section~6 we overview further tasks to be performed and compare our results with previous work on heavy baryon form factors. Finally, in Section~7 we conclude.

\section{Recall of the sum rules in the meson case.} \hspace*{\parindent}
In the case of mesons, the general SR obtained from the OPE can be written in the compact way \cite{4r}
\beq
\label{8e}
L_{Hadrons} (w_i, w_f, w_{if}) = R_{OPE} (w_i, w_f, w_{if})
\eeq

\noi where the l.h.s. is the sum over the intermediate $D$ states, while the r.h.s. is the OPE counterpart. This expression writes, in the heavy quark limit \cite{4r}~:
\bea
\label{9e}
&&\sum_{D=P,V}\sum_n Tr \left [ \overline{B}_f (v_f) \Gamma_f D^{(n)}(v')\right ] \ Tr \left [ \overline{D}^{(n)} (v') \Gamma_i B_i(v_i)\right ] \xi^{(n)}(w_i) \xi^{(n)}(w_f)\nn \\
&&+ \hbox{ Other excited states} = - 2 \xi (w_{if}) \ Tr \left [ \overline{B}_f (v_f) \Gamma_f P'_+ \Gamma_i B_i(v_i)\right ]
\eea

\noi where
\beq
\label{10e}
w_i = v_i \cdot v' \qquad w_f = v_f \cdot v' \qquad w_{if} = v_i \cdot v_f
\eeq

\noi $P'_+ = \displaystyle{{1 + {/\hskip - 2 truemm v}' \over 2}}$ is the positive energy projector on the intermediate $c$ quark. We assume that the IW functions are real and the $\overline{B}$ meson is the pseudoscalar ground state $(j^P, J^P) = \left ( {1 \over 2}^-, 0^-\right )$, $j$ is the angular momentum of the light cloud and $J$ the spin of the bound state. The heavy quark currents considered in the preceding expression are 
\beq
\label{11e}
\overline{h}_{v'}\Gamma_i h_{v_i} \qquad\qquad  \overline{h}_{v_f}\Gamma_f h_{v'}
\eeq

\noi In (9) $B(v)$, $D(v)$ are the $4 \times 4$ matrices representing the $\overline{B}$, $D$ states \cite{7r}, and $\overline{B} = \gamma^0B^+\gamma^0$ denotes the Dirac conjugate matrix. The domain for the variables $(w_i, w_f, w_{if})$ is \cite{4r}~:
$$w_i \geq 1 \qquad\qquad w_f \geq 1$$
\beq
\label{12e}
w_iw_f - \sqrt{(w_i^2-1)(w_f^2-1)} \leq w_{if} \leq w_iw_f + \sqrt{(w_i^2 - 1) (w_f^2 -1)}
\eeq

\noi Taking $w_i = w_f = w$, the domain becomes
\beq
\label{13e}
w \geq 1 \qquad\qquad 1 \leq w_{if} \leq 2w^2-1
\eeq

In \cite{4r} the following SR were established. Taking $\Gamma_i = {/\hskip-2 truemm v}_i$ and $\Gamma_f = {/\hskip-2 truemm v}_f$ one finds the so-called Vector SR
$$(w+1)^2 \sum_{L\geq 0} {L+1 \over 2L+1}\ S_L (w,w_{if}) \sum_n \left [ \tau_{L+1/2}^{(L)(n)}(w)\right ]^2+ \sum_{L\geq 1}  S_L (w,w_{if}) \sum_n \left [ \tau_{L-1/2}^{(L)(n)}(w)\right ]^2$$
\beq
\label{14e}
= \left ( 1 + 2w + w_{if}\right ) \xi (w_{if})
\eeq

\noi and for $\Gamma_i = {/\hskip-2 truemm v}_i\gamma_5$ and $\Gamma_f = {/\hskip-2 truemm v}_f\gamma_5$ one finds the Axial SR

 $$\sum_{L\geq 0}  S_{L+1} (w,w_{if}) \sum_n \left [ \tau_{L+1/2}^{(L)(n)}(w)\right ]^2 +(w-1)^2 \sum_{L\geq 1}  {L \over 2L-1} S_{L-1} (w,w_{if}) \sum_n \left [ \tau_{L-1/2}^{(L)(n)}(w)\right ]^2$$
\beq
\label{15e} 
=-  \left ( 1 - 2w + w_{if}\right ) \xi (w_{if})
\eeq

\noi In the precedent equations the IW functions $\tau_{L\pm 1/2}^{(L)(n)}(w)$ correspond to the transitions ${1 \over 2}^- \to\left (L \pm {1 \over 2} \right )^P$,  $P = (-1)^{L+1}$,  and the function $S_{L} (w,w_{if})$ is given by the Legendre polynomial
\beq
\label{16e}
S_L(w, w_{if}) = \sum_{0 \leq k \leq L/2} C_{L,k}\left ( w^2-1\right )^{2k} \left (w^2 - w_{if}\right )^{L-2k}
\eeq

\noi with 
\beq
\label{17e}
C_{L,k} = (-1)^k {(L!)^2 \over (2L)!}\ {(2L-2k)! \over k!(L-k)!(L-2k)!}
\eeq
\vskip 5 truemm

Differentiating $n$ times both SR (\ref{14e}), (\ref{15e}) with respect to $w$ and $w_{if}$ and going to the border of the domain (\ref{13e}) $w_{if} = w = 1$, one gets the bounds (\ref{1e})-(\ref{5e}). \par

To be complete in this recall of the meson case, let us remind that, on the other hand, Uraltsev \cite{18r} did propose a special limit of HQET, namely the so-called BPS limit, that implies
\beq
\label{18e}
\rho^2 = {3 \over 4}
\eeq

\noi among other interesting consequences for subleading quantities. In \cite{19r} it was demonstrated, using the above SR, that if the slope reaches its lower bound  (\ref{1e}), as happens in the BPS limit, then all derivatives reach their lower bounds  (\ref{5e}), and then the IW function is completely fixed, namely 
\beq
\label{19e}
\xi (w) = \left ( {2 \over w+1}\right )^{3/2}
\eeq

\section{Sum rules for the baryon case.} \hspace*{\parindent} 
As explained in \cite{8r}, the following fields can be written for the $J = j \pm {1 \over 2}$ baryons, where $j$ is the spin of the light cloud.\par

One defines the tensor-spinor (we change Falk's notation $A^{\mu_1 \cdots \mu_j}$ by $\varepsilon^{\mu_1 \cdots \mu_j}$ to underline what is technically common with the meson case)
\beq
\label{20e}
\psi^{\mu_1 \cdots \mu_j} = \varepsilon^{\mu_1 \cdots \mu_j}\ u_h
\eeq

\noi where $\varepsilon^{\mu_1 \cdots \mu_j}$ is symmetric, and the following transversity and tracelessness conditions are fulfilled
\bea
\label{21e}
&v_{\mu_k}\ \varepsilon^{\mu_1 \cdots \mu_j}= 0 &\qquad (k = 1, \cdots j) \nn \\
&g_{\mu_i \mu_k}\ \varepsilon^{\mu_1 \cdots \mu_j}= 0 &\qquad (i,k = 1, \cdots j)\qquad (i\not= k)
\eea

\noi Then, there are two baryon fields corresponding to $J = j \pm {1 \over 2}$

$\psi^{\mu_1 \cdots \mu_j} _{j-1/2}$
\beq
\label{22e}
= {1 \over 2j+1} \left [ \left ( \gamma^{\mu_1} + v^{\mu_1} \right ) \gamma_{\nu_1}\ g_{\nu_2}^{\mu_2} \cdots g_{\nu_j}^{\mu_j} + \cdots + g_{\nu_1}^{\mu_1} \cdots g_{\nu_{j-1}}^{\mu_{j-1}} \left ( \gamma^{\mu_j} + v^{\mu_j} \right )  \gamma_{\nu_j} \right ] \varepsilon^{\nu_1 \cdots \nu_j}\ u_h
\eeq

$\psi^{\mu_1 \cdots \mu_j} _{j+1/2} = \Big \{ g_{\nu_1}^{\mu_1} \cdots g_{\nu_j}^{\mu_j}$
\beq
\label{23e}
- {1 \over 2j+1} \left [ \left ( \gamma^{\mu_1} + v^{\mu_1} \right ) \gamma_{\nu_1}\ g_{\nu_2}^{\mu_2} \cdots g_{\nu_j}^{\mu_j} + \cdots + g_{\nu_1}^{\mu_1} \cdots g_{\nu_{j-1}}^{\mu_{j-1}} \left ( \gamma^{\mu_j} + v^{\mu_j} \right )  \gamma_{\nu_j} \right ] \Big \} \varepsilon^{\nu_1 \cdots \nu_j}\ u_h
\eeq

\noi where $u_h$ is the spin ${1 \over 2}$ field of the heavy quark. \par

It follows, from (\ref{21e}),
\bea
\label{24e}
&v_{\mu_k}\ \psi^{\mu_1 \cdots \mu_j}_{j\pm 1/2} = 0 &\qquad (k = 1, \cdots j) \nn \\
&g_{\mu_i \mu_k}\ \psi^{\mu_1 \cdots \mu_j}_{j\pm 1/2}= 0 &\qquad (i,k = 1, \cdots j)\qquad (i\not= k)
\eea

The tensor (23) is the generalization to all $j$ of the Rarita-Schwinger vector-spinor field, that satisfies another condition,
\beq
\label{25e}
\gamma_{\mu_k}\ \psi^{\mu_1 \cdots \mu_j}_{j + 1/2} = 0 \qquad (k = 1, \cdots j) 
\eeq

\noi The matrix elements of a heavy quark current
\beq
\label{26e}
\overline{h}_{v'}\Gamma h_{v} 
\eeq

\noi for the transition $j \pm {1 \over 2} \to j' \pm {1 \over 2}$ writes \cite{7r}
\beq
\label{27e}
<H'_{j'\pm 1/2} (v')|J(q)|H_{j\pm 1/2}(v)>\ = \overline{\psi}{'}_{j' \pm 1/2}^{\nu_1 \cdots \nu_{j'}} \ \Gamma \ \psi^{\mu_1 \cdots \mu_j}_{j\pm 1/2} \ \zeta_{\nu_1 \cdots \nu_{j'}; \mu_1 \cdots \mu_j} 
\eeq

\noi where $j' \geq j$ is assumed and the tensor $\zeta_{\nu_1 \cdots \nu_{j'}; \mu_1 \cdots \mu_j}$ is given by the expression
\bea
\label{28e}
&&\zeta_{\nu_1 \cdots \nu_{j'}; \mu_1 \cdots \mu_j} = (-1)^j (v' - v)_{\nu_{j+1}} \cdots (v' - v)_{\nu_{j'}} \Big [ C_0^{(j',j)} (w) g_{\nu_1 \mu_1} \cdots g_{\nu_j \mu_j}\nn \\
&& + C_1^{(j',j)}(w) (v'-v)_{\nu_1}  (v'-v)_{\mu_1}  g_{\nu_2 \mu_2} \cdots g_{\nu_j \mu_j} + \cdots \nn \\
&& + C_j^{(j',j)}(w) (v'-v)_{\nu_1}  (v'-v)_{\mu_1} \cdots (v'-v)_{\nu_j}(v'-v)_{\mu_j}\Big ] 
\eea

\noi If $H(v)$ is the ground state $\Lambda_b$ (angular momentum of the light quarks $j = 0$), one has a much simpler expression
\beq
\label{29e}
<H'_{j\pm 1/2} (v')|J(q)|\Lambda_b(v)>\ = \overline{\psi}{'}_{j' \pm 1/2}^{\nu_1 \cdots \nu_{j'}}\ \Gamma \ u_h  \ \zeta_{\nu_1 \cdots \nu_{j'}} 
\eeq

\noi with 
\beq
\label{30e}
\zeta_{\nu_1 \cdots \nu_{j'}} = (v'-v)_{\nu_1}  \cdots (v'-v)_{\nu_{j'}} C_0^{(j',0)} (w)
\eeq

\noi and one is left, therefore, with a unique class of IW functions $C_0^{(j',0)} (w)$ for the transition $0 \to j'$. From now on we adopt the notation
\beq
\label{31e}
 \tau_{j'}^{(n)}(w)= C_0^{(j',0)(n)} (w)
\eeq

\noi where $n$ denotes a possible radial quantum number of the final baryon $H^{'(n)}$. \par

Considering the non-forward transition
\beq
\label{32e}
\Lambda_b(v_i) \to \Lambda_c^{(n)}(v') \to \Lambda_b (v_f)
\eeq

\noi and general currents
\beq
\label{33e}
\overline{h}_{v'}\Gamma_i h_{v_i} \qquad {\rm and} \qquad  \overline{h}_{v_f}\Gamma_f h_{v'}
\eeq

\noi We find, in an analogous way to the meson SR,
\bea
\label{34e}
&&\sum_n \sum_L \tau_{L}^{(n)}(w_i) \ \tau_{L}^{(n)}(w_f)\ v_{f\mu_1} \cdots v_{f\mu_L}\ v_{i\nu_1} \cdots v_{i\nu_L}\nn \\
&&\left ( \overline{u}_{h_f} \overline{\Gamma}_f \psi{'}^{\mu_1 \cdots \mu_L}_{L+1/2} \ \overline{\psi}{'}^{\nu_1 \cdots \nu_L}_{L+1/2} \Gamma_i u_{h_i} + \overline{u}_{h_f}\overline{\Gamma}_f \psi{'}^{\mu_1 \cdots \mu_L}_{L-1/2} \ \overline{\psi}{'}^{\nu_1 \cdots \nu_L}_{L-1/2} \Gamma_i u_{h_i} \right ) \nn \\
&&= \xi_\Lambda (w_{if}) \left [ \overline{u}_{h_f} \overline{\Gamma}_f \Lambda '_+ \Gamma_i u_{h_i}\right ]
\eea

\noi (where $ \overline{\Gamma} = \gamma^0 \Gamma^+ \gamma^0$) and we denote the elastic IW function by $\xi_\Lambda (w)$, identical to 
\beq
\label{35e}
\xi_\Lambda (w) =  \tau_{0}^{(0)}(w)
\eeq

\noi Let us underline that in the case of the SR $\Lambda_b (v_i) \to \Lambda_c^{(n)}(v') \to \Lambda_b (v_f)$, only intermediate states $ \Lambda_c^{(n)}$ with isospin $I=0$ can contribute and, in general, all $j^P = L^P$, $P = (-1)^L$, contribute. Hence the notation $\tau_{L}^{(n)}(w)$ for the transition IW functions.

\subsection{Vector sum rule.} \hspace*{\parindent}
Adopting vector currents aligned along the intermediate four-velocity $v'$
\beq
\label{36e}
\Gamma_i = \overline{\Gamma}_i = {/\hskip - 2 truemm v}'  \qquad\qquad \Gamma_f = \overline{\Gamma}_f = {/\hskip - 2 truemm v}'  
\eeq

\noi and denoting the tensor
\beq
\label{37e}
T^{\rho_1 \cdots \rho_L,\sigma_1 \cdots \sigma_L} = \sum_{\lambda} \varepsilon{'}^{(\lambda ) \star \rho_1 \cdots \rho_L} \ \varepsilon{'}^{(\lambda ) \sigma_1 \cdots \sigma_L} 
\eeq

\noi where the sum is carried over the $2L+1$ polarizations $\lambda = -L, \cdots + L$, we find, using the symmetry properties of the tensor (37),
\bea
\label{38e}
&&\sum_n \sum_L \tau_{L}^{(n)}(w_i) \ \tau_{L}^{(n)}(w_f)\ T^{\rho_1 \cdots \rho_L,\sigma_1 \cdots \sigma_L} \Big \{ v_{f\rho_1} \cdots v_{f\rho_L}\ v_{i\sigma_1} \cdots v_{i\sigma_L} \left [  \overline{u}_{h_f}  \Lambda '_+ u_{h_i}\right ] \nn \\
&&-{L \over 2L+1} \  v_{f\rho_2} \cdots v_{f\rho_L}\ v_{i\sigma_2} \cdots v_{i\sigma_L}\nn \\
&&\left ( \left ( w_f + 1\right ) v_{i\sigma_1} \left [ \overline{u}_{h_f} \gamma_{\rho_1} \Lambda '_+ u_{h_i}\right ] + \left ( w_i + 1\right ) v_{f\rho_1} \left [ \overline{u}_{h_f} \Lambda '_+ \gamma_{\sigma_1} u_{h_i}\right ] \right ) \nn \\
&&+ \ 2\left ( {L \over 2L+1}\right )^2 \left ( w_i + 1\right ) \left ( w_f + 1\right )   v_{f\rho_2} \cdots v_{f\rho_L}\ v_{i\sigma_2} \cdots v_{i\sigma_L} \left [ \overline{u}_{h_f} \gamma_{\rho_1} \Lambda '_+  \gamma_{\sigma_1} u_{h_i}\right ] \Big \}\nn \\
&&= \xi_\Lambda (w_{if}) \left [ \overline{u}_{h_f} \Lambda '_+  u_{h_i}\right ]
\eea
\noi and $\Lambda '_+$ denotes the positive energy projector on the intermediate spinors $\Lambda '_+ = \sum\limits_{s'} u_{v'}^{(s')} \overline{u}_{v'}^{(s')}$. \par

In some aspects, this expression seems simpler than the one for the meson case. However, now the SR depends not only on the quantity \cite{4r}
\beq
\label{39e}
S_L(w_i,w_f,w_{if}) = T^{\rho_1 \cdots \rho_L,\sigma_1 \cdots \sigma_L} \ v_{f\rho_1} \cdots v_{f\rho_L}\ v_{i\sigma_1} \cdots v_{i\sigma_L}
\eeq

\noi but also on the quantities
\bea
\label{40e}
&&T_L^{(1)}(w_i, w_f, w_{if}) = T^{\rho_1 \cdots \rho_L,\sigma_1 \cdots \sigma_L} \ v_{f\rho_2} \cdots v_{f\rho_L}\ v_{i\sigma_1} \cdots v_{i\sigma_L}  \left [ \overline{u}_{h_f} \gamma_{\rho_1} \Lambda '_+ u_{h_i}\right ] \nn\\
&&T_L^{(2)}(w_i, w_f, w_{if}) = T^{\rho_1 \cdots \rho_L,\sigma_1 \cdots \sigma_L} \ v_{f\rho_1} \cdots v_{f\rho_L}\ v_{i\sigma_2} \cdots v_{i\sigma_L}  \left [ \overline{u}_{h_f} \Lambda '_+ \gamma_{\sigma _1}  u_{h_i}\right ] \nn\\
&&U_L(w_i, w_f, w_{if}) = T^{\rho_1 \cdots \rho_L,\sigma_1 \cdots \sigma_L} \ v_{f\rho_2} \cdots v_{f\rho_L}\ v_{i\sigma_2} \cdots v_{i\sigma_L}  \left [ \overline{u}_{h_f} \gamma_{\rho_1} \Lambda '_+ \gamma_{\sigma_1} u_{h_i}\right ] \nn\\
\eea

\noi To compute (\ref{40e}) we need the tensor
\beq
\label{41e}
S_L^{\rho_1, \sigma_1} = {\partial^2 \over \partial  {{v_f}_{\rho_1}} \partial {{v_i}_{\sigma_1}}}\ S_L(w_i,w_f,w_{if})  
\eeq

\noi A calculation using the expression \cite{4r}
 \beq
\label{42e}
S_L(w_i,w_f,w_{if})   = \sum_{0 \leq k \leq {L \over 2}} C_{L,k}\left  ( w_i^2 - 1\right )^k \left ( w_f^2 - 1 \right )^k \left ( w_i w_f - w_{if}\right )^{L-2k}
\eeq

\noi that reduces to (\ref{16e}) for $w_i = w_f = w$, and $C_{L,k}$ is given by (\ref{17e}), yields 
\bea
\label{43e}
&&S_L^{\rho_1, \sigma_1} = {1 \over L^2} \sum_{0 \leq k \leq {L \over 2}} C_{L,k}\left  ( w_i^2 - 1\right )^k \left ( w_f^2 - 1 \right )^k \left ( w_i w_f - w_{if}\right )^{L-2k}\nn \\
&&\Big [ (L-2k) \left ( w_i w_f - w_{if}\right )^{-1}\left (  v{'}^{\rho_1} v{'}^{\sigma_1}- g^{\rho_1\sigma_1}\right ) \nn \\
&&+ \ 2k(L-2k)  \left  ( w_i^2 - 1\right )^{-1} \left ( w_i w_f - w_{if}\right )^{-1} \left (  w_i v{'}^{\rho_1} - v_i^{\rho_1}\right )  \left (  w_i v{'}^{\sigma_1} - v_i^{\sigma_1}\right )\nn \\
&&+\  (L-2k)  (L-2k-1)  \left ( w_i w_f - w_{if}\right )^{-2} \left (  w_i v{'}^{\rho_1} - v_i^{\rho_1}\right )  \left (  w_f v{'}^{\sigma_1} - v_f^{\sigma_1}\right )\nn \\
&&+ \ 4k^2 \left  ( w_i^2 - 1\right )^{-1} \left  ( w_f^2 - 1\right )^{-1} \left (  w_f v{'}^{\rho_1} - v_f^{\rho_1}\right )  \left (  w_i v{'}^{\sigma_1} - v_i^{\sigma_1}\right )\nn \\
&&+ \ 2k(L-2k)\left  ( w_f^2 - 1\right )^{-1} \left ( w_i w_f - w_{if}\right )^{-1}  \left (  w_f v{'}^{\rho_1} - v_f^{\rho_1}\right )  \left (  w_f v{'}^{\sigma_1} - v_f^{\sigma_1}\right )\Big ]
\eea

\noi A straightforward but lengthy calculation gives
\bea
\label{44e}
&&T_L^{(1)}(w_i, w_f, w_{if}) = {1 \over L} \sum_{0 \leq k \leq {L \over 2}} C_{L,k}\left  ( w_i^2 - 1\right )^k \left ( w_f^2 - 1 \right )^k \left ( w_i w_f - w_{if}\right )^{L-2k}\nn \\
&&\Big \{ \left [  \left ( w_i w_f - w_{if}\right )^{-1}\left  ( w_f + 1\right ) (L-2k) + \left  ( w_i + 1\right )^{-1} 2k \right ]  \left [  \overline{u}_{h_f}  \Lambda '_+ u_{h_i}\right ]\nn \\
&&-\left ( w_i w_f - w_{if}\right )^{-1}\left  ( w_f + 1\right ) (L-2k) \left [  \overline{u}_{h_f}  u_{h_i}\right ]\Big \} \\
&&\nn \\
&&T_L^{(2)}(w_i, w_f, w_{if}) = T_L^{(1)}(w_f, w_i, w_{if}) \\
\label{45e}
&& \nn \\
&&U_L(w_i, w_f, w_{if}) ={1 \over L^2} \sum_{0 \leq k \leq {L \over 2}} C_{L,k}\left  ( w_i^2 - 1\right )^k \left ( w_f^2 - 1 \right )^k \left ( w_i w_f - w_{if}\right )^{L-2k}\nn \\
&&\Big \{ \Big [ (L-2k)  \left ( w_i w_f - w_{if}\right )^{-1} (3+4k) + (L-2k)(L-2k-1)  \left ( w_i w_f - w_{if}\right )^{-2}  \nn \\
&&\left ( w_i w_f + w_i + w_f - 1 - 2w_{if}\right )+4k^2 \left ( w_i +1\right )^{-1} \left ( w_f +1\right )^{-1}\Big ] \left [  \overline{u}_{h_f}  \Lambda '_+ u_{h_i}\right ] \nn \\
&&- \left ( w_i w_f - w_{if}\right )^{-1}(L-2k)(2L+1)\Big ]\left [  \overline{u}_{h_f}  u_{h_i}\right ] \Big \}
\label{46e}
\eea

Using these expressions in the SR (\ref{38e}) we find that the coefficient of the bilinear $[  \overline{u}_{h_f}  u_{h_i}]$ {\it vanishes identically}, and the coefficient of $[ \overline{u}_{h_f}  \Lambda '_+ u_{h_i}]$, non-vanishing in general, gives the SR
\bea
\label{47e}
&&\sum_n \sum_L \tau_{L}^{(n)}(w_i) \ \tau_{L}^{(n)}(w_f)   \sum_{0 \leq k \leq {L \over 2}} C_{L,k}\left  ( w_i^2 - 1\right )^k \left ( w_f^2 - 1 \right )^k\Big \{  \left ( w_i w_f - w_{if}\right )^{L-2k} \nn \\
&&-{2 \over 2L+1} \left [ (L-2k)    \left ( w_i +1\right ) \left ( w_f +1\right ) \left ( w_i w_f - w_{if}\right )^{L-2k-1} + 2k \left ( w_i w_f - w_{if}\right )^{L-2k}\right ] \nn \\
&&+ \  {2 \over (2L+1)^2} \Big [ (L-2k) (3 + 4k)\left ( w_i +1\right ) \left ( w_f +1\right ) \left ( w_i w_f - w_{if}\right )^{L-2k-1}\nn \\ 
&&+\  (L-2k)(L-2k-1) \left ( w_i +1\right ) \left ( w_f +1\right ) \left ( w_i w_f + w_i + w_f - 1 - 2w_{if}\right )\nn \\
&&\left ( w_i w_f - w_{if}\right )^{L-2k-2}+ \ 4k^2 \left ( w_i w_f - w_{if}\right )^{L-2k}\Big ] \Big \} = \xi_\Lambda (w_{if})
\eea

\noi Recall that in the sum of the l.h.s. the IW functions 
\bea
\label{48bise}
\xi_\Lambda^{(n)}(w) = \tau_{0}^{(n)}(w) \qquad\qquad \xi_\Lambda(w) = \tau_{0}^{(0)}(w)
\eea		
also appear, and $C_{L,k}$ is given by (\ref{17e}). \par

As we did for mesons, for our purpose it is enough to take $w_i = w_f =w$, that gives the simpler expression
\bea
\label{48e}
&&\sum_{L \geq 0} \ \sum_{n \geq 0} \left [ \tau_{L}^{(n)}(w) \right ]^2  \sum_{0 \leq k \leq {L \over 2}} C_{L,k}\left  ( w^2 - 1\right )^{2k} \Big \{ \left ( w^2 - w_{if} \right )^{L-2k}  \nn \\
&&-{2 \over 2L+1} \left  [ (L-2k)    \left ( w +1\right )^2 \left ( w^2 -w_{if} \right )^{L-2k-1}+ 2k \left ( w^2 - w_{if}\right )^{L-2k}\right ] \nn \\
&&+\ {2 \over (2L+1)^2} \Big [ (L-2k) (3 + 4k)\left ( w +1\right )^2 \left ( w^2- w_{if}\right )^{L-2k-1}\nn \\ 
&&+\  (L-2k)(L-2k-1) \left ( w +1\right )^2 \left ( w^2 +2w- 1- 2w_{if}\right )^{L-2k-2}\nn \\
&&+\  4k^2 \left ( w^2 - w_{if}\right )^{L-2k}\Big ] \Big \} = \xi_\Lambda (w_{if})
\eea

\subsection{Axial sum rule.} \hspace*{\parindent} 
Following the analogy with the calculation in the meson case, if we take, instead of the vector currents (\ref{36e}), the axial currents aligned also along the intermediate four-velocity $v'$
\beq
\label{49e}
\Gamma_i = \overline{\Gamma}_i = {/\hskip - 2 truemm v}' \ \gamma_5  \qquad\qquad \Gamma_f = \overline{\Gamma}_f = {/\hskip - 2 truemm v}'  \ \gamma_5
\eeq

\noi we obtain, instead of (\ref{48e}), the expression

\bea
\label{50e}
&&\sum_{L \geq 0} \ \sum_{n \geq 0} \left [ \tau_{L}^{(n)}(w) \right ]^2  \sum_{0 \leq k \leq {L \over 2}} C_{L,k}\left  ( w^2 - 1\right )^{2k} \Big \{ \left ( w^2 - w_{if} \right )^{L-2k}  \nn \\
&&-{2 \over 2L+1} \left  [ (L-2k)    \left ( w - 1\right )^2 \left ( w^2 -w_{if} \right )^{L-2k-1}+ 2k \left ( w^2 - w_{if}\right )^{L-2k}\right ] \nn \\
&&+{2 \over (2L+1)^2} \Big [ (L-2k) (L+2+ 2k)\left ( w -1\right )^2 \left ( w^2- w_{if}\right )^{L-2k-1}\nn \\ 
&&- (L-2k)(L-2k-1) \left ( w -1\right )^2 \left ( 2w + w_{if} + 1\right ) \left ( w^2 - w_{if}\right )^{L-2k-2}\nn \\
&&+ \ 4k^2 \left ( w^2 - w_{if}\right )^{L-2k}\Big ] \Big \} = \xi_\Lambda (w_{if})
\eea

\section{Bounds on the derivatives of the Isgur-Wise function.} \hspace*{\parindent}
We will now exploit the Vector SR (\ref{48e}), by computing its derivatives and going to the frontier of the domain (13) $w \to 1$, $w_{if} \to 1$ 
 \beq
\label{51e}
\left ( {d^{p+q} \over dw_{if}^p dw^q} \right )_{w_{if}=w=1}
\eeq

\noi For arbitrary $p$ and $q=0$, one finds
\beq
\label{52e}
 \xi_\Lambda^{(p)}(1) = (-1)^p p! \sum_{n\geq 0}  \left [ \tau_{p}^{(n)}(1) \right ]^2
\eeq

\noi We recover therefore the result of Isgur et al. [8] for the slope (6) and generalize it for any derivative, giving
\beq
\label{53e}
(-1)^p \ \xi_\Lambda^{(p)}(1) \geq 0
\eeq

\noi that demonstrates that the IW function $\xi_\Lambda (w)$ is an alternate series in powers of $(w-1)$. \par

We find the same result (\ref{53e}) from the Axial SR using (50)-(51).

\section{Improved bound on the curvature.} 
\subsection{Vector sum rule.} \hspace*{\parindent}
Moreover, an improved bound can be found on the curvature, similar to the one found in the meson case (\ref{3e}).\par

To obtain it, let us consider the Vector SR (\ref{48e}) and (\ref{51e}) for different values for $p$, $q$ satisfying $p+q \leq 2$. \par

For $p = q = 0$ we obtain $[ \xi_\Lambda (1) ]^2 = \xi_\Lambda (1)$, i.e. $1=1$. For both cases $p = 1$, $q=0$ or $p=0$, $q=1$ we obtain
 \beq
\label{54e}
 \xi '_\Lambda (1) = - \sum_{n\geq 0}  \left [ \tau_{1}^{(n)}(1) \right ]^2
\eeq

\noi i.e. eq. (\ref{6e}). For $p = 2$, $q=0$ we get equation (\ref{52e}) for $p=2$, 
\beq
\label{55e}
 \xi ''_\Lambda (1) = 2 \sum_{n\geq 0}  \left [ \tau_{2}^{(n)}(1) \right ]^2
\eeq

\noi while for $p = 1$, $q=1$ one gets
 \beq
\label{56e}
2 \sum_{n\geq 0}  \left [ \tau_{2}^{(n)}(1) \right ]^2 + \sum_{n\geq 0}  \tau_{1}^{(n)}(1)\  \tau{'}_{1}^{(n)}(1) = 0
\eeq

\noi and finally for $p=0$, $q=2$,
 \bea
\label{57e}
&&\sum_{n\geq 0}  \left [ \tau_{1}^{(n)}(1) \right ]^2 + {8 \over 3} \sum_{n\geq 0}  \left [ \tau_{2}^{(n)}(1) \right ]^2 + \sum_{n\geq 0}  \left [  \xi {'}^{(n)}_\Lambda (1)\right ]^2 \nn \\
&&+\ 4  \sum_{n\geq 0}  \tau_{1}^{(n)}(1)\  \tau{ '}_{1}^{(n)}(1) +  \xi ''_\Lambda (1) = 1
\eea

\noi where we have used the notation (\ref{48bise}). Eliminating the unknown $\sum\limits_{n\geq 0}  \tau_{1}^{(n)}(1)  \tau{'}_{1}^{(n)}(1)$ between eqs. (\ref{56e}) and (\ref{57e}), and using (\ref{54e}) we obtain finally for the curvature,
 \beq
\label{58e}
\sigma_\Lambda^2 = \xi ''_\Lambda (1) = {3 \over 5} \left \{  \rho_\Lambda^2 + (\rho_\Lambda^2)^2 + \sum_{n\not= 0} \left [  \xi {'}^{(n)} (1) \right ]^2 \right \}
\eeq

\noi that implies the improved bound
 \beq
\label{59e}
\sigma_\Lambda^2 = \xi ''_\Lambda (1) \geq {3 \over 5} \left [ \rho_\Lambda^2 + (\rho_\Lambda^2)^2\right ]
\eeq

\subsection{Axial sum rule.} \hspace*{\parindent} 
Let us now consider the Axial SR (\ref{50e}) and use (\ref{51e}) for different values for $p$, $q$ satisfying $p+q \leq 3$.\par

For $p = q = 0$ we obtain a trivial result. For ($p=1$, $q=0$), ($p=0$, $q=1$), ($p=2$, $q=0$), ($p=1$, $q=1$) and ($p=0$, $q=2$) we get respectively the same equations (\ref{54e})-(\ref{57e}) as for the Vector SR. \par

Notice that in the meson case we got different SR for the vector and the axial currents. This corresponds to the fact underlined above that, in the meson case, in the SR we have contributions that for a given $L$, the light cloud has two possible values $j^P = \left ( L \pm {1 \over 2}\right )^P$, $P = (-1)^{L+1}$. \par

For the baryon transition $\Lambda_b \to \Lambda_c$, since the two spectator quarks have total spin and isospin $S = I = 0$, for a given $L$ we have a single type of intermediate states, with $j^P = L^P$. This explains why we obtain less information in the baryon case than in the meson case for the elastic IW function.\par
Although not providing new information, the consideration of the present case with the axial current remains interesting as a check of the results found for the vector current.

 \section{Prospects and comparison with previous work.} \hspace*{\parindent}
 Some work should be pursued on this subject. On the one hand, within our approach, bounds on higher derivatives may be obtained and could be useful.\par
 
On the other hand, one should include the radiative corrections to the slope and curvature of $\xi_\Lambda (w)$ and $1/m_Q$ corrections within Heavy Quark Effective Theory, as well as the Wilson coefficients that make the matching with the physical form factors. This program was performed in the meson case by Dorsten \cite{20r}, and should be carried out in the baryon case. \par 
These improvements should be accomplished in order to make a realistic comparison with experiment and with models for the different $\Lambda_b \to \Lambda_c \ell \overline{\nu}_{\ell}$ form factors at finite mass.\par
However, within rigorous QCD methods, there is one theoretical scheme that can be directly confronted with our bound, namely the IW function computed in the heavy quark and large $N_c$ limits \cite{10r}. Notice that our bounds hold in the heavy quark limit at the actual value $N_c = 3$.\par
The result at large $N_c$ \cite{10r} is a simple exponential form

\beq
\label{59bise}
\xi_\Lambda (w) = exp \left [ - \rho_\Lambda^2 (w-1)\right ]
\eeq

\noi with 

\beq
\label{60bise}
\rho_\Lambda^2 = \lambda N^{3/2}\qquad\qquad\lambda = O(1)
\eeq
The bound for the curvature (\ref{59e}) implies for the slope of the exponential form (\ref{59bise}) 

\beq
\label{61bise}
\rho_\Lambda^2 \geq 3/2
\eeq
that, from (\ref{60bise}), is trivially satisfied in the large $N_c$ limit. However, the phenomenological guess (3.8) from \cite{10r}, $ \rho_\Lambda^2 = 1.3 $, slightly violates the bound. But we should keep in mind that a guess obtained from the large $N_c$ limit not necessarily should satisfy a constraint obtained in the physical case  $N_c = 3$.\par
The other work on baryon form factors $\Lambda_b \to \Lambda_c \ell \overline{\nu}_{\ell}$ that is based on QCD is the dispersive approach of Boyd et al. \cite{2bisr,3bisr}, that uses, {\it at finite mass}, crossing symmetry, dispersion relations and pertubative QCD evaluated far from the physical resonances. This method, that has received a number of improvements in the meson case, gives in principle a model independent description of the various form factors in terms of a finite number of parameters. In the most favorable case, it allows for baryons to describe one of the helicity amplitudes in terms of two unknown constants \cite{3bisr}. Notice that, unlike appealing to the crossed channel $\Lambda_b \overline{\Lambda}_c$, our results hold directly in the semileptonic region.\par
Since the dispersive approach of Boyd et al. is formulated directly at finite mass, the comparison with the present work would require to carefully obtain the relation between the IW function  $\xi_\Lambda (w)$ and the physical form factors along the lines exposed at the beginning of the present Section. On the other hand, ref. \cite{3bisr} does not give, unlike in the meson case (cf. the relation between the curvature and the slope, formula (7.7)), simple formulae for the baryon form factors, and one would need to carefully extract the results for baryons from the general formalism. Because of these two reasons, we postpone to future work the comparison of the results of the present paper with the ones of \cite{3bisr}.

 \section{Conclusion.} \hspace*{\parindent}
 In conclusion, from the OPE sum rules obtained in the heavy quark limit of QCD using the non-forward amplitude, we get an expansion of the elastic Isgur-Wise function  $\xi_\Lambda (w)$ for the process $\Lambda_b \to \Lambda_c \ell \overline{\nu}_{\ell}$, up to order $(w-1)^2$,
\beq
\label{60e}
\xi_\Lambda (w) = 1 - \rho_\Lambda^2 (w-1) + {\sigma_\Lambda^2 \over 2} (w-1)^2 + O\left [ (w-1)^3\right ]
\eeq

\noi with the constraints
\beq
\label{61e}
\rho_\Lambda^2 \geq 0 \qquad\qquad\qquad \sigma_\Lambda^2 \geq {3 \over 5} \left [\rho_\Lambda^2 + \left ( \rho_\Lambda^2 \right )^2\right ] 
\eeq

\noi While the first inequality (\ref{61e}) for the slope was known \cite{8r}, the second one is new, the main result of this paper. On the other hand, we have demonstrated that $\xi_\Lambda (w)$ is an alternate series in powers of $(w-1)$. \par

Notice the important point that if the slope is not small, of $O(1)$, owing to the available phase space the curvature will have a measurable effect, and will be important in the extrapolation of the differential rate at $w \to 1$.  \par

On the experimental side, hopefully the LHC-b program would provide information on the shape of this function, and help to further constrain the CKM matrix element $V_{cb}$. Let us recall that there is a small tension between the exclusive $\overline{B} \to D^{(*)}\ell \overline{\nu}_{\ell}$ and the inclusive $\overline{B} \to X_c \ell \overline{\nu}_{\ell}$ determinations of $V_{cb}$, although the error of the former determination is rather large. \par

On the theoretical side, one should include the radiative corrections to the slope and curvature of $\xi_\Lambda (w)$ and $1/m_Q$ corrections, as well as the Wilson coefficients that make the matching with the physical form factors. This would allow to compare with future data and with other theoretical or phenomenological schemes of baryon form factors at finite mass. Also, once these necessary improvements are realized, any future fit to the differential distribution of the process $\Lambda_b \to \Lambda_c \ell \overline{\nu}_{\ell}$ using the Isgur-Wise function $\xi_\Lambda (w)$ should take into account our constraints.

\end{document}